\documentclass{article}
\usepackage{amssymb}
\usepackage{amsfonts}
\usepackage{amsmath}
\usepackage{epsfig}
\usepackage{graphicx}

\begin{document}

\title{On three-body effects and nuclear fusion}
\author{R. Vilela Mendes\thanks{%
e-mail: rvilela.mendes@gmail.com; rvmendes@ciencias.ulisboa.pt;
https://label2.tecnico.ulisboa.pt/vilela/}}
\date{ }
\maketitle

\begin{abstract}
Confined to small regions, quantum systems exhibit electronic and structural
properties different from their free space behavior. In Coulomb 3-body
problems, configurations of close proximity of identically charged particles
are classically unstable. They however exist as excited quantum states in
confined systems. Quantum control of such states might be useful to induce
nuclear fusion. The difficult nature of the quantum control of these {\it %
quantum collision states} is discussed here, as well as a possible solution
using a two-step nonunitary evolution process.
\end{abstract}

\section{Confined systems, three-body effects and nuclear reactions}

This introductory section addresses the question of Coulomb systems on
confined environments,\ with a particular view towards its relevance for
small scale fusion energy applications. A few personal notes are also
included which, if not relevant for the objective validity of the results
that are presented, might nevertheless be of interest to a reader concerned
with the way ideas are developed.

Quantum systems in confined environments, for example molecules embedded in
other material lattices, display chemical reaction properties different from
the free space behavior \cite{MolContainer}. Could the same be true for
nuclear reactions and could they be useful?

One such event could be the ill-famed Fleischmann-Pons effect. In March 1989
Fleischmann and Pons \cite{Fleisch} claimed to have obtained, at room
temperature, nuclear fusion with deuterium atoms absorbed by electrolysis
into a palladium electrode. Similar claims were made a few months later by
another group \cite{Jones}. These claims were met with justified disbelief,
if nothing else on grounds of the energy scales involved. How could a
shielding effect in the solid-state environment, presumably on the order of
a few $eV$'s, be sufficient to overcome the Coulomb barrier? Furthermore,
the results could not be reproduced by experiments under carefully
controlled conditions (\cite{Berlinguette} and references therein).

Shortly after the Fleischman-Pons announcement, a meeting was organized in
Marseille, at the CPT-CNRS\footnote{%
Centre de Physique Th\'{e}orique - Centre Nationale de la Recherche
Scientifique - France}, about this phenomenon. The big star at the meeting
was the work of Leggett and Baym \cite{Leggett} which showed that, under
equilibrium conditions, the rate of tunneling to a $r=0$ separation of the
deuterons was rigorously bounded above by the value calculated by the
Born-Oppenheimer potential. Much too small to induce any meaningful fusion
rates. In fact, if the effective repulsion at short distances of the
deuterons were to be much reduced by the solid-state environment, then one
would also expect a much increased binding affinity of $\alpha -$particles
to the metal, which is not observed. The work of Leggett and Baym definitely
excluded any equilibrium, ground state or even low-lying effect.

Particularly annoyed by the Fleischmann-Pons effect were the plasma
physicists, involved in the magnetic confinement program. They had no reason
to be. Even if controlled lattice-confined nuclear reactions were achieved,
they would probably be useful for small scale applications, not to replace
the planned large scale fusion reactors, to follow the DEMO prototype, as
base load sources of electricity. More serious competitors, in the future,
for large scale energy installations, might be deep-drilling geothermics or
thorium reactors.

Clearly not being a ground state or a low-lying excitation, could the
Fleischmann-Pons effect be a low probability phenomena, associated to some
unusual (chaotic?) system configuration of two nucleons plus the crystal
electronic sea? If a real effect, their erratic and irreproducible nature
seemed to point in that direction. Together with Samuel Eleut\'{e}rio a
back-of-the-envelop ergodic calculation was performed showing that, in a
chaotic system, the three-body system of two deuterons and one electron had
a non-negligiible probability for quasi-collision events. That is, events
with the interparticle distances to be as small as desired. In addition, the
reaction channels of the three-body event favoured a chain of reactions that
suppressed the production of neutrons. The results were included in two
preprints \cite{IFM-9-89} \cite{IFM-10-89} which were sent to Nature. They
were instantly returned by the editor with the quote "{\it there is no point
is speculating about non-existent effects}". An improved version of the
ergodic calculation was later published \cite{VilelaMPLB1}, slightly
disguised as an exercise in classical statistical mechanics.

Although the three-body interpretation seemed a reasonable possibility, to
convert the confining environment into an ergodic system without destroying
the crystal itself was not a very practical perspective. Therefore it seemed
to me, at the time, that, even if real, the Fleischmann-Pons effect might
never have a practical application as an energy production mechanism. My
interest in the subject dwindled. It was only revived when, at a meeting in
Karpacz, I learned from Semenov the existence of the quantum scar effect 
\cite{Heller} \cite{Berry}. In the mathematical physics literature, the
notion of scar state became rather restrictive, but what it means in
practice is the appearance of a stationary well-defined wave function
located around some unstable classical orbit. Being interested in quantum
complexity \cite{Karpacz} and quantum control, this suggested an interesting
possibility. In the classical case it is always very difficult to make use
of unstable orbits, however, if these orbits have a scarred quantum
counterpart, these states might be addressed by resonant excitation at the
appropriate energy. Scar states\ were indeed found to exist in charged
Coulomb problems \cite{Vilela-scar1}, with a particularly interesting case
corresponding to configurations related to unstable saddle points of the
potential ({\it saddle scars }\cite{Vilela-scar2}).

Scar-like states where positive particles are in close proximity ({\it %
quantum collision states} as defined below) are indeed found in Coulomb
systems. The current availability of laser pulses of diverse shapes and time
scales, as used in the control of molecular dynamics, would suggest that the
quantum collision states might be easy to address by quantum control with
electromagnetic radiation. It turns out that even in simple theoretical
models \cite{Protocol} \cite{Vilela-maxsym} that is not the case. In this
paper one discusses the reason for this fact and a possible solution using a
nonunitary\footnote{%
Not continuously unitary} two-step control mechanism.

\section{Quantum collision states in a three-body Coulomb system}

Here one considers a Coulomb system with two heavy positively charged
particles and one light negatively charged one. Choose coordinates in the
center of mass of the two heavy particles with the $z-$axis along the line
connecting the heavy particles. Hence, the coordinates of the three-body
system are $\left( \widetilde{R},\widetilde{\rho },\theta ,\phi \right) $,
the Cartesian coordinates of the heavy particles being $\left( 0,0,\frac{%
\widetilde{R}}{2}\right) $ and $\left( 0,0,-\frac{\widetilde{R}}{2}\right) $
and the spherical coordinates of the light negative charge $\left( 
\widetilde{\rho },\theta ,\phi \right) $. The Hamiltonian of the system is 
\begin{equation}
\widetilde{H}_{0}=-8\frac{\hslash ^{2}}{M}\frac{\partial ^{2}}{\partial 
\widetilde{R}^{2}}-\frac{\hslash ^{2}}{2m}\left\{ \frac{1}{\widetilde{\rho }%
^{2}}\frac{\partial }{\partial \widetilde{\rho }}\left( \widetilde{\rho }^{2}%
\frac{\partial }{\partial \widetilde{\rho }}\right) +\frac{1}{\widetilde{%
\rho }^{2}\sin \theta }\frac{\partial }{\partial \theta }\left( \sin \theta 
\frac{\partial }{\partial \theta }\right) +\frac{1}{\widetilde{\rho }%
^{2}\sin ^{2}\theta }\frac{\partial ^{2}}{\partial \phi ^{2}}\right\}
+V\left( \widetilde{R},\widetilde{\rho },\theta \right)   \label{QC1}
\end{equation}%
with%
\begin{equation}
\widetilde{V}\left( \widetilde{R},\widetilde{\rho },\theta \right) =\frac{%
Z^{2}e^{2}}{4\pi \epsilon _{0}}\frac{1}{\widetilde{R}}-\frac{Zqe^{2}}{4\pi
\epsilon _{0}}\left\{ \frac{1}{\sqrt{\frac{\widetilde{R}^{2}}{4}+\widetilde{%
\rho }^{2}-\widetilde{R}\widetilde{\rho }\cos \theta }}+\frac{1}{\sqrt{\frac{%
\widetilde{R}^{2}}{4}+\widetilde{\rho }^{2}+\widetilde{R}\widetilde{\rho }%
\cos \theta }}\right\}   \label{QC2}
\end{equation}%
$M$ and $m$ being the masses of the positive and negative charges and $Z$
and $q$ their charges.

For the system to be useful both for nuclear and molecular studies, it is
convenient to scale down the variables and work with dimensionless
quantities, $H_{0},R,\rho $. 
\begin{eqnarray}
R &=&G^{2}\widetilde{R}  \nonumber \\
\rho &=&G^{2}\widetilde{\rho }  \nonumber \\
H_{0} &=&\frac{2m}{\hslash ^{2}G^{4}}\widetilde{H}_{0}  \label{QC3}
\end{eqnarray}%
with \ $G^{2}=\frac{Ze^{2}m}{2\pi \epsilon _{0}\hslash ^{2}}$. Then%
\begin{eqnarray}
H_{0} &=&-8\mu \frac{\partial ^{2}}{\partial R^{2}}-\frac{\hslash ^{2}}{2m}%
\left\{ \frac{1}{\rho ^{2}}\frac{\partial }{\partial \rho }\left( \rho ^{2}%
\frac{\partial }{\partial \rho }\right) +\frac{1}{\rho ^{2}\sin \theta }%
\frac{\partial }{\partial \theta }\left( \sin \theta \frac{\partial }{%
\partial \theta }\right) +\frac{1}{\rho ^{2}\sin ^{2}\theta }\frac{\partial
^{2}}{\partial \phi ^{2}}\right\} +V\left( R,\rho ,\theta \right)  \nonumber
\\
V &=&\frac{Z}{R}-q\left\{ \frac{1}{\sqrt{\frac{R^{2}}{4}+\rho ^{2}-R\rho
\cos \theta }}+\frac{1}{\sqrt{\frac{R^{2}}{4}+\rho ^{2}+R\rho \cos \theta }}%
\right\}  \label{QC4}
\end{eqnarray}%
with $\mu =\frac{m}{M}$.

One defines as $\eta -${\it quantum collision states }the states for which
the wave function norm $\left\vert \Psi \left( 0,0,\theta ,\phi \right)
\right\vert ^{2}>\eta $ for some $\theta ,\phi $. To be of practical
importance one should require $\eta $ to be sufficiently large. Because the
potential in (\ref{QC4}) has extensive attractive regions (Fig.\ref%
{Pot_average}), one might naively expect these quantum collision states to
easily occur. However, this not so because the kinetic energy of the light
particle dominates the Hamiltonian and low energy quantum collision states
do not occur. Worse than that, their peculiar nature renders them isolated
states and difficult to excite by quantum control.

\begin{figure}[htb]
\centering
\includegraphics[width=0.75\textwidth]{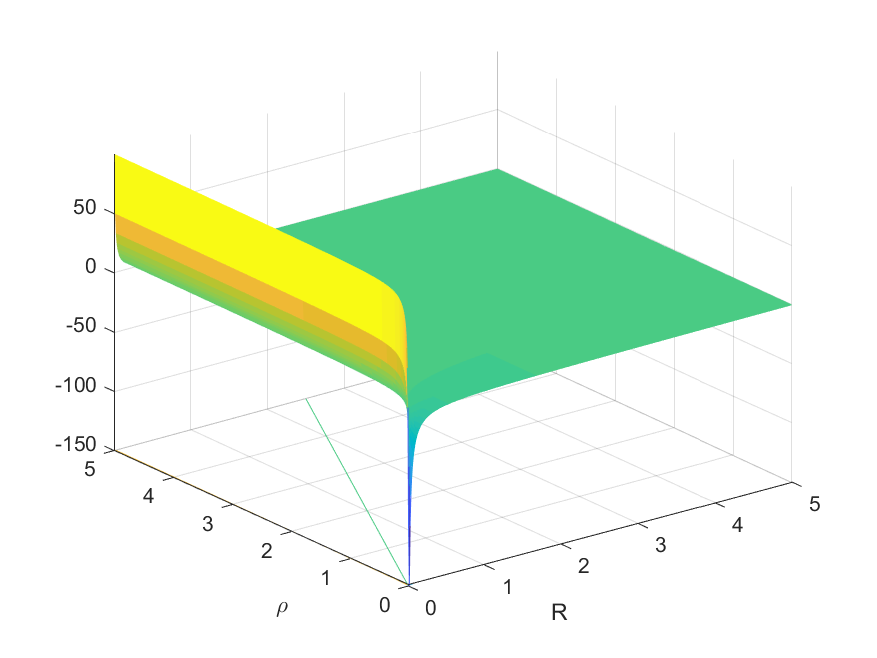}
\caption{$\theta -$averaged
potential ($Z=q=1)$}
\label{Pot_average}
\end{figure}

Figs.\ref{Pot_average} shows the potential $V\left( R,\rho ,\theta \right) $%
, averaged over $\theta $%
\[
\overline{V}\left( R,\rho \right) =\frac{1}{2}\int_{-1}^{1}d\left( \cos
\theta \right) V\left( R,\rho ,\theta \right) 
\]%

Notice that all the region below
the line $\rho =\frac{R}{2}$ is an attractive region. However, for small $R$
values the potential is strongly repulsive, except if also $\rho $ is very
small.

Exploration of high excited states in the confined system is better carried
out by implementing a discretization of the operators, derivatives and
Laplacians, on a lattice, and performing the diagonalization of the
resulting discretized Hamiltonian. To be relevant for a specific physics
problem, that is, to be able to relate the dimensionless results to physical
quantities, the size of the volume where the system is confined should be
adjusted to the value of $G^{2}$ (see below). For the moment let us consider
the three-body system to be confined in a spherical volume of
(dimensionless) radius $5$.

Consider the wave function $\Psi \left( R,\rho ,\theta ,\phi \right) $
factorized as%
\[
\Psi \left( R,\rho ,\theta ,\phi \right) =\Psi \left( R,\rho ,\theta \right)
e^{il\phi } 
\]%
To consider values $l\neq 0$ only adds a scalar factor to the Hamiltonian.
Therefore here only the $l=0$ case is considered.

\begin{figure}[htb]
\centering
\includegraphics[width=0.75\textwidth]{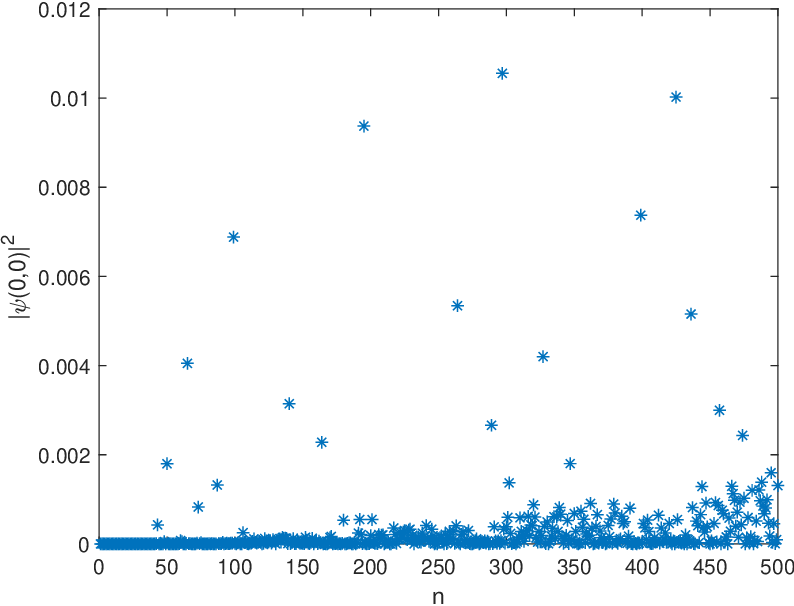}
\caption{Quantum collision probabilities}
\label{Quantum_colli}
\end{figure}

Performing the numerical diagonalization of the Hamiltonian, the whole
spectrum is obtained and one finds the quantum collision probabilities,%
\[
p_{QC}^{(i)}=\int d\left( \cos \theta \right) \left\vert \Psi _{i}\left(
0,0,\theta \right) \right\vert ^{2}
\]%
This is displayed in Fig.\ref{Quantum_colli}

\begin{figure}[htb]
\centering
\includegraphics[width=0.75\textwidth]{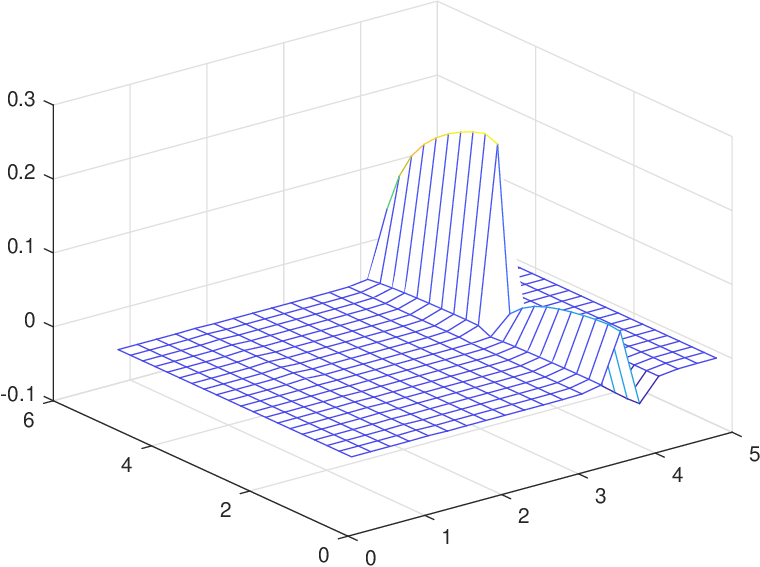}
\caption{Ground state wave function}
\label{Ground_state}
\end{figure}

\begin{figure}[htb]
\centering
\includegraphics[width=0.75\textwidth]{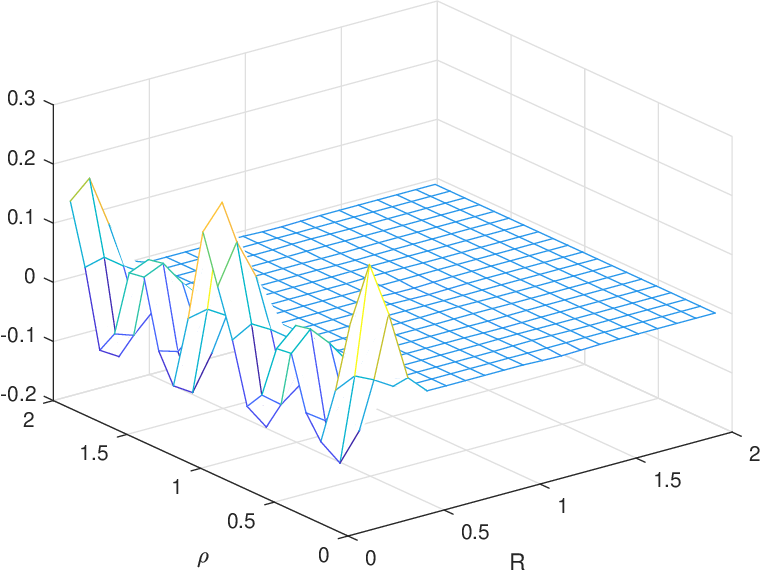}
\caption{A quantum collision state}
\label{QColliWave}
\end{figure}

\begin{figure}[htb]
\centering
\includegraphics[width=0.75\textwidth]{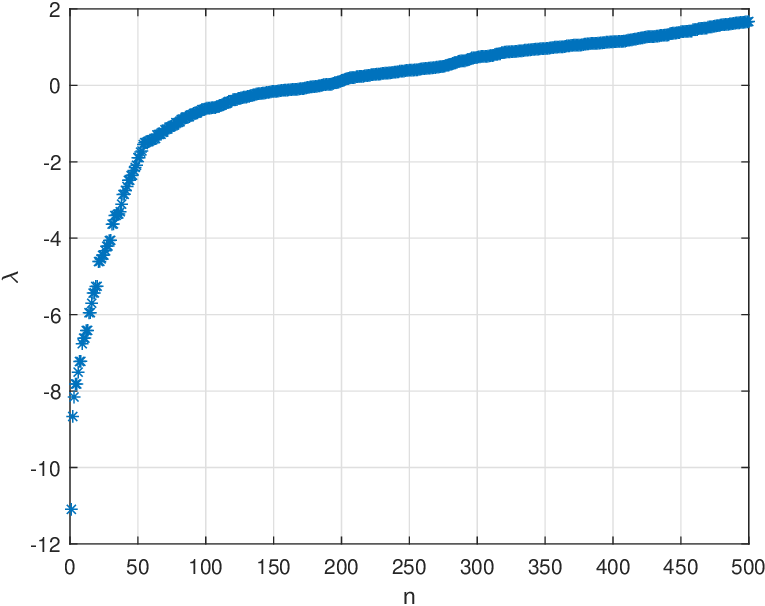}
\caption{The spectrum}
\label{Spectrum}
\end{figure}

One sees clearly that these
states are located high in the spectrum and they are isolated states. Figs.%
\ref{Ground_state} and \ref{QColliWave} show the image of the wave functions
at $\theta =\frac{\pi }{2}$ for the ground state and a for typical quantum
collision state. On the other hand the
eigenvalues of the spectrum are displayed in Fig.\ref{Spectrum}.

To quantify in physical units these results, the value of $G^{2}=%
\frac{Ze^{2}m}{2\pi \epsilon _{0}\hslash ^{2}}$ must be computed. For
definiteness one considers here $m$ to be the electron mass, $M$ the
deuteron mass and $Z=q=1$. Then%
\begin{eqnarray*}
\frac{G^{2}}{Z} &=&\frac{e^{2}m}{2\pi \epsilon _{0}\hslash ^{2}}%
=3.77945\times 10^{10}meter^{-1} \\
\frac{Z}{G^{2}} &=&2.6459\times 10^{-11}meter=0.26459\mathring{A} \\
\mu  &=&\frac{m}{M}=2.724447\times 10^{-4}
\end{eqnarray*}%
Physical lengths, $\widetilde{R}$ and $\widetilde{\rho }$, are related to
the dimensionless quantities $R$ and $\rho $ by%
\begin{eqnarray*}
\widetilde{R} &=&R\times \frac{Z}{G^{2}} \\
\widetilde{\rho } &=&\rho \times \frac{Z}{G^{2}}
\end{eqnarray*}%
As a consequence, a spherical volume of radius $5$ corresponds roughly to
the volume of an octahedral cage in a face-centered palladium crystal and
radius $4$ to a similar cage in nickel. For the conversion of energy
quantities one has%
\[
\widetilde{H}_{0}=\frac{\hslash ^{2}G^{4}}{2mZ^{2}}H_{0}=H_{0}\times
0.87195\times 10^{-17}J=H_{0}\times 54.42277eV
\]%
Another important conversion factor is the time scaling which is important
when considering the time-dependent Schr\"{o}dinger equation%
\begin{eqnarray*}
i\hslash \frac{\partial }{\partial \widetilde{t}}\Psi  &=&\widetilde{H}%
_{0}\Psi =\frac{\hslash ^{2}G^{4}}{2mZ^{2}}H_{0}\Psi  \\
i\frac{\partial }{\partial \left( \frac{\hslash G^{4}}{2mZ^{2}}\widetilde{t}%
\right) }\Psi  &=&i\frac{\partial }{\partial t}\Psi =H_{0}\Psi 
\end{eqnarray*}%
Therefore%
\[
\widetilde{t}=\frac{2mZ^{2}}{\hslash G^{4}}t=1.20944\times 10^{-17}s
\]

With these conversion factors one may now estimate the energy of the quantum
collision states. For the state shown in Fig.\ref{QColliWave} it is roughly $%
10$ units above the ground state, which corresponds to an excitation energy
of $544$ eV. This would be the energy of a photon of wavelength $2.278\times
10^{-9}m$ in the low X-ray region. A few other isolated states with lower $%
p_{QC}$ probabilities appear but no lower than $5$ (dimensionless)
excitation units\footnote{%
These are energies at least one hundred times higher than thermal
excitations. Therefore one should no expect the quantum collision states to
be excited by thermal fluctuations. Given the scale of energies needed to
excite them, the quantum collision state might not be at the origin of the
cold fusion events. When observed, they might also arise from a mixture of
exceptional causes. For example, when deuterium is absorbed by hydrogen
storage materials, there is an expansion of the crystal lattice and cracks
are expected to occur. Strong electric fields may occur in the cracks
accelerating the deuterons to nuclear fusion energies. This is consistent
with the cases where protuberances and craters in cold fusion samples are
observed, these being often the sites where unexpected elements appear in
high local concentrations.
\par
On the other hand in hot spots caused by the (aggressive) electrolysis
process, instead of an ordered structure, one might have a hot soup of
electrons and deuterons and then, in this ergodic situation, it is known
that three body collisions DeD have a small but non-negligible probability.
The occurrence of both situations, of course, will very much depend on the
material structure of the samples. Being probably due to a mixture of
exceptional events, it is therefore natural for spontaneous cold fusion
events to be hardly reproducible in any controllable way.}.

The energy gap between the ground state and the quantum collision states
makes it clear that any practical utility of such states would rely on
devising a reliable reproducible way to excite such states. Their isolated
nature in an otherwise non-quantum-collisional spectrum is the first
difficulty. Even if one finds out exactly the value of the excitation
energy, there is no guarantee that shining an electromagnetic radiation of
such energy on the confining crystal would excite the quantum collision
states. A whole ladder of vibrational levels may be excited, as is well
known when designing fluorescent and phosphorescent materials. However, the
situation is even worse than that, as explained later.

\begin{figure}[htb]
\centering
\includegraphics[width=0.75\textwidth]{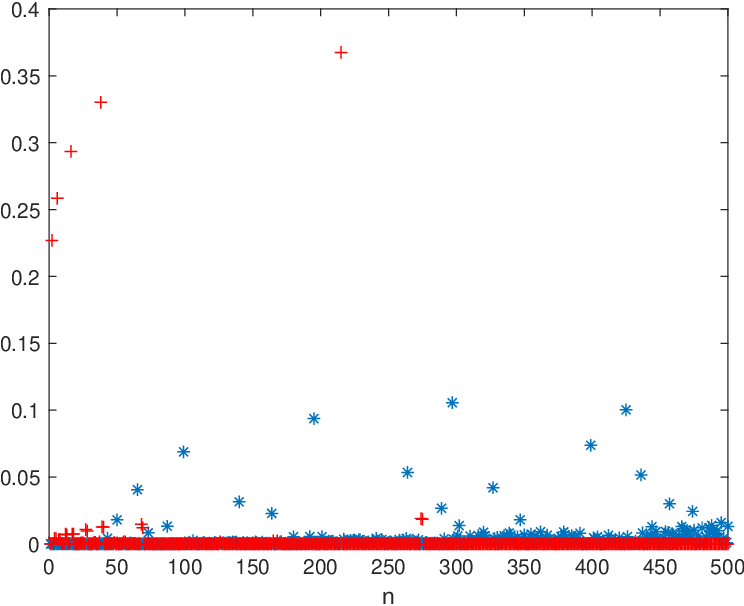}
\caption{Paramagnetic transition
probabilities (+) compared with the positions (*)\ of the quantum collision
states}
\label{Transition}
\end{figure}

Let us denote by $P_{+}^{(1)}$ and $P_{+}^{(2)}$ the momenta of the positive
charges and $p_{-}$ the momentum of the negative one. The Hamiltonian of the
system in interaction with an electromagnetic field is%
\[
H_{em}=\frac{\left( P_{+}^{(1)}-eA\right) ^{2}}{2M}+\frac{\left(
P_{+}^{(2)}-eA\right) ^{2}}{2M}+\frac{\left( p_{-}+eA\right) ^{2}}{2m}+e\Phi
\left( \frac{R}{2}\right) +e\Phi \left( -\frac{R}{2}\right) -e\Phi \left(
\rho ,\theta ,\phi \right) +V\left( R,\rho ,\theta \right) 
\]%
In the $\nabla \cdot A=0$ gauge%
\begin{eqnarray*}
H_{em} &=&H_{0}+H_{1} \\
H_{1} &=&-\frac{eA\cdot P_{+}^{(1)}}{M}-\frac{eA\cdot P_{+}^{(2)}}{M}+\frac{%
eA\cdot p_{-}}{m}+e^{2}A^{2}\left( \frac{1}{M}+\frac{1}{2m}\right) +e\Phi
\left( \frac{R}{2}\right) +e\Phi \left( -\frac{R}{2}\right) -e\Phi \left(
\rho ,\theta ,\phi \right) 
\end{eqnarray*}%
the first three terms being the paramagnetic terms and the fourth the
diamagnetic one. In the center of mass of positively charged particles the
first two terms cancel each other and, for a linearly polarized
monochromatic plane wave,%
\begin{eqnarray*}
\Phi  &=&0 \\
A &=&A_{0}\overrightarrow{\epsilon }\cos \left( k\cdot r-\omega t\right) 
\end{eqnarray*}%
the paramagnetic term is simply%
\[
H_{1}^{P}=\frac{eA_{0}\overrightarrow{\epsilon }\cdot p_{-}}{2m}\left(
e^{i\left( k\cdot r-\omega t\right) }+e^{-i\left( k\cdot r-\omega t\right)
}\right) 
\]%
In first order, the paramagnetic transition probability from the ground
state $\left\vert 0\right\rangle $ to an excited state $\left\vert
s\right\rangle $ is%
\[
p_{0\rightarrow s}^{P}\left( t\right) =\frac{t^{2}}{\hbar ^{2}}\frac{%
e^{2}\left\vert A_{0}\right\vert ^{2}}{4m^{2}}\left\vert \left\langle
s\left\vert \overrightarrow{\epsilon }\cdot p_{-}\exp \left( ik\cdot
r\right) \right\vert 0\right\rangle \right\vert ^{2}\text{sinc}^{2}\left[
\left( \omega -\frac{\Delta E_{s}}{\hbar }\right) t/2\right] 
\]%
$\Delta E_{s}$ being the excitation energy of the state $\left\vert
s\right\rangle $. Because of the sinc$^{2}$ term, $\omega $ should be close
to the resonant frequency $\frac{\Delta E_{s}}{\hbar }$. However, that is
not sufficient, the matrix element $\left\vert \left\langle s\left\vert 
\overrightarrow{\epsilon }\cdot p_{-}\exp \left( ik\cdot r\right)
\right\vert 0\right\rangle \right\vert $ should also be large. For
wavelengths larger than the size of the system this is dominated by $%
\left\vert \left\langle s\left\vert \overrightarrow{\epsilon }\cdot
p_{-}\right\vert 0\right\rangle \right\vert $. From the spectrum previously
obtained, this matrix element has been computed, with the Laplacian in
spherical coordinates and the polarization $\overrightarrow{\epsilon }$ in a
generic position. The result (red crosses) is shown in Fig.\ref{Transition}

One sees that the states for which the paramagnetic matrix element is large
are never the quantum collision states (blue *). In fact this was to be
expected, because from the relation $\left[ r,H\right] =\frac{i\hbar p}{m}$
one knows that the relevant matrix element is simply the dipole moment and
naturally quantum collision states are bound to have a negligible dipole
momentum.

\begin{figure}[htb]
\centering
\includegraphics[width=0.75\textwidth]{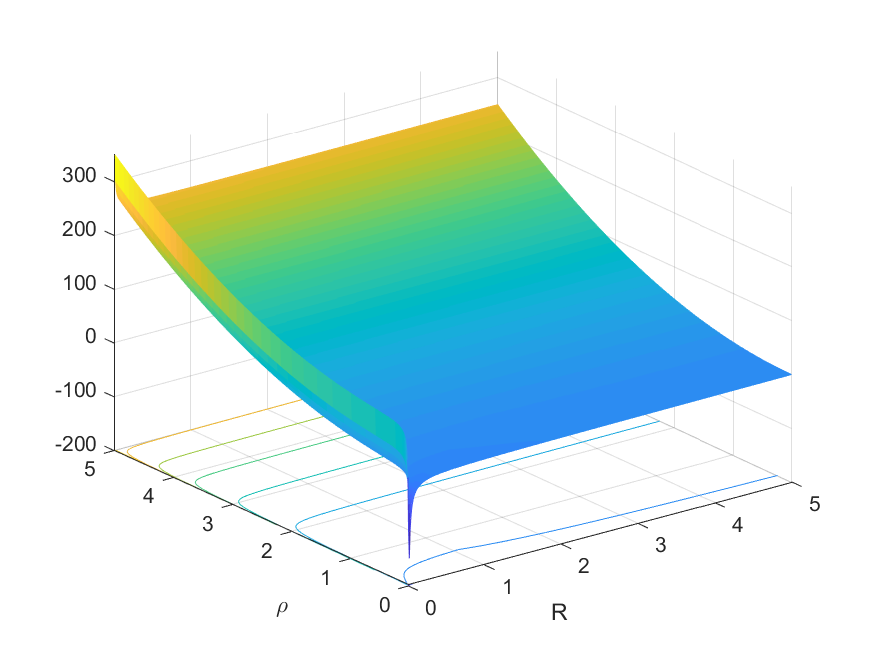}
\caption{$\theta -$averaged potential on a strong magnetic field}
\label{Pot_average_mag}
\end{figure}

On the other hand, diamagnetic terms introduce an harmonic interaction on
the plane orthogonal to the magnetic field. For a constant magnetic field%
\[
A=\frac{1}{2}\left( \overrightarrow{r}\times \overrightarrow{B}\right) 
\]%
one obtains%
\[
\frac{e^{2}}{2}A^{2}=\frac{e^{2}}{4}\left( r^{2}B^{2}-\left( \overrightarrow{%
r}\cdot \overrightarrow{B}\right) ^{2}\right) 
\]%
Fig.\ref{Pot_average_mag} shows the effective averaged potential for a
strong constant magnetic field. One sees that the harmonic interaction
favours wave functions with smaller average $\rho $ values, but the
potential and the Hamiltonian spectrum are not qualitatively different.
Therefore, also the diamagnetic interaction cannot provide a simple sure
control mechanism to excite, in a steady reproducible manner, the quantum
collision states.

Of course, the negative paramagnetic result, previously obtained with the $%
H_{1}^{P}$ matrix elements, is only a first order result. However one should
not expect very different behavior from higher order corrections. The non
controllability of the quantum collision states by direct electromagnetic
interaction, described here is consistent with previous results, where a
probably difficult to achieve scheme with an objective Lyapunov function was
attempted for a maximally symmetric 3-body system \cite{Vilela-maxsym}.
There it was found that although the adjustable control achieves some
decreasing of the interparticle distance, absolute control of the collision
states is not achieved. This essentially means that the trajectory in
Hilbert space of states, that is induced by the electromagnetic interaction,
starting from the ground state, never reaches the goal collision state. This
is a familiar situation where the goal state cannot be reached by unitary
control. One possible solution is the scheme of non-unitary control first
proposed in \cite{Manko-Strocchi} and later developed by Mandilara and Clark %
\cite{Mandilara} and other authors \cite{Rabitz}. The idea is: at a certain
point in the unitary trajectory, a measurement is performed, making the
system jump to a different quantum trajectory which might then reach the
goal state. In the present context this might mean that instead of trying to
control the goal state by a resonant frequency equal to $\frac{\Delta E_{s}}{%
\hbar }$, one excites the system to some intermediate state, from which, by
a subsequent electromagnetic interaction, the goal state might be reached.
Having the full spectrum numerically computed before, this situation may be
explored here.

\begin{figure}[htb]
\centering
\includegraphics[width=0.75\textwidth]{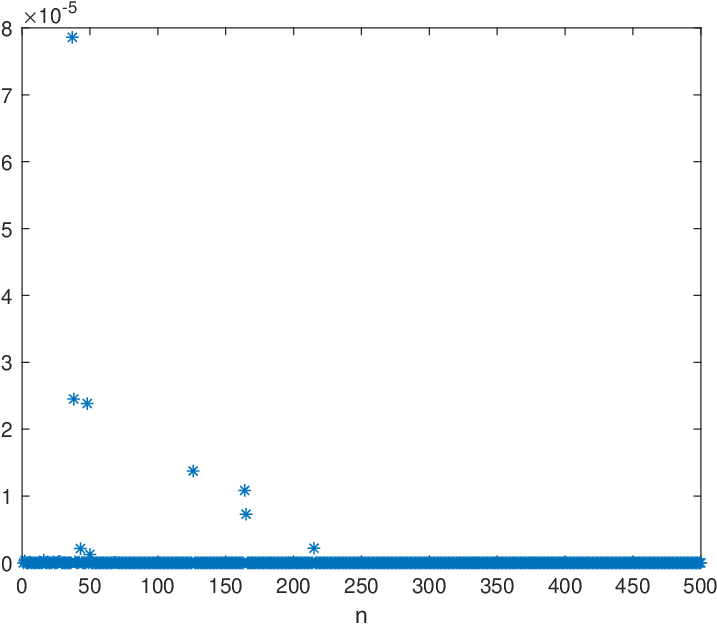}
\caption{Estimate of the two-step transtion probabilities}
\label{CTR}
\end{figure}

\begin{figure}[htb]
\centering
\includegraphics[width=0.75\textwidth]{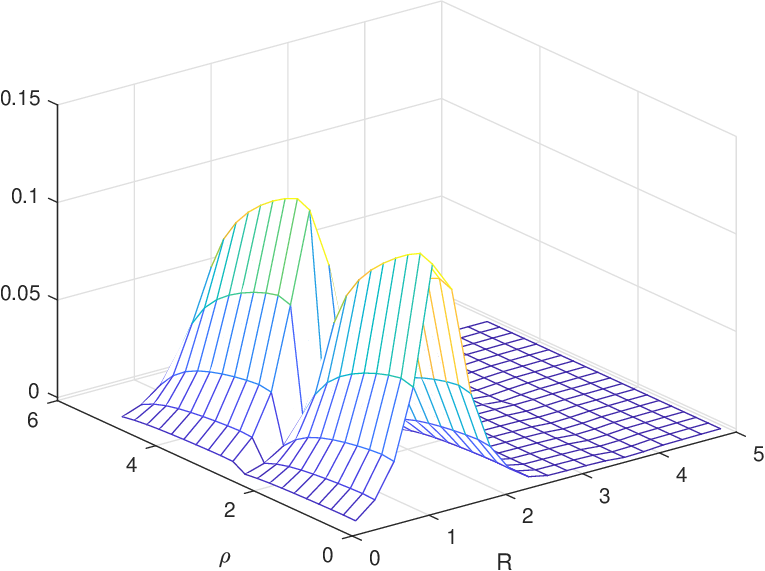}
\caption{An intermediate state for the
two-step excitation of the quantum collision state}
\label{IntermState}
\end{figure}

For this purpose one computes both the matrix element $\left\vert
\left\langle n\left\vert \overrightarrow{\epsilon }\cdot p_{-}\right\vert
0\right\rangle \right\vert $ from the ground state to all other states as
well as the corresponding matrix elements $\left\vert \left\langle
n\left\vert \overrightarrow{\epsilon }\cdot p_{-}\right\vert s\right\rangle
\right\vert $ from the quantum collision state shown in Fig.\ref{QColliWave}%
. Assuming that in each case the appropriate resonant frequencies $\frac{%
\Delta E_{0\rightarrow n}}{\hbar }$ and $\frac{\Delta E_{n\rightarrow s}}{%
\hbar }$ are used, multiplying these matrix elements, one obtains an
estimate of a two-step transition from the ground state to the quantum
collision state. That is, the first transition $\left\vert 0\right\rangle
\rightarrow \left\vert n\right\rangle $ would prepare the system in the
state $\left\vert n\right\rangle $, acting like a measurement, with the
second transition reaching the goal $\left\vert s\right\rangle $. The result
of the calculation is shown in Fig.\ref{CTR}. One sees that there are a few
states below and above the energy level of the quantum collision state for
which the two-step transition probability is non-negligible. For the quantum
collision state shown in Fig.\ref{QColliWave} one such state would be for
example the one shown in Fig.\ref{IntermState}. The two-step control becomes
possible because, when acted upon by the $\overrightarrow{\epsilon }\cdot
p_{-}$ operator, this intermediate state has non-negligible overlaps with
both the ground state and the quantum collision one.

In this example the (dimensionless) excitation energy of the quantum
collision state would be $10.4754$ with the intermediate state located at
energy $7.7865\,$. With the conversion factor $\frac{2.278\times 10^{-8}}{H}%
m $, found before, this means that instead of a resonant pulse with
wavelength $2.175\times 10^{-9}m$, one should use a preparation pulse of
wavelength $2.926\times 10^{-9}m$ followed by one with $8.472\times 10^{-9}m$%
.

Of course, one could also imagine a unitary evolution with a continuously
adjusted wavelength. However this sounds unpractical.

In conclusion: quantum control of the quantum collision states, by a
two-step process, seems possible although, of course, to be implemented in
the laboratory it would require a very accurate calculation or measurement
of the excitation spectrum. Alternatively, one might implement an automatic
random exploration of pairs of different frequency pulses in the low X-ray
and high ultraviolet range.

\end{document}